\documentclass[prd,superscriptaddress,showpacs,onecolumn,showkeys]{revtex4}
\usepackage{amsmath}
\usepackage{amssymb}
\usepackage{graphicx}
\usepackage{listings}
\usepackage{palatino}
\usepackage[colorinlistoftodos]{todonotes}
\usepackage{mathpazo}
\usepackage{mathtools}
\usepackage[colorlinks=true,citecolor=blue,linkcolor=blue,anchorcolor=green, anchorcolor=green,urlcolor=blue]{hyperref}
\usepackage{amsfonts}
\usepackage{longtable}
\usepackage{amsthm}
\usepackage{xcolor}
\usepackage{eurosym}
\usepackage{amsfonts}
\usepackage{calrsfs}

\DeclareFontEncoding{LGR}{}{}

\ProvideTextCommand{\~}{LGR}[1]{\char126#1}
\newcommand{\bq}{\begin{equation}}
\newcommand{\eq}{\end{equation}}
\newcommand{\bqn}{\begin{eqnarray}}
\newcommand{\eqn}{\end{eqnarray}}

\begin{document}

\title{Quasinormal Modes of a Schwarzschild Black Hole Immersed in an
Electromagnetic Universe}
\author{Ali \"{O}vg\"{u}n}
\email{ali.ovgun@pucv.cl}
\affiliation{Instituto de F\'{\i}sica, Pontificia Universidad Cat\'olica de Valpara\'{\i}%
so, Casilla 4950, Valpara\'{\i}so, Chile.}
\affiliation{Physics Department, Arts and Sciences Faculty, Eastern Mediterranean
University, Famagusta, North Cyprus via Mersin 10, Turkey.}
\affiliation{School of Natural Sciences, Institute for Advanced Study, 1 Einstein Drive Princeton, NJ 08540, USA.}

\author{\.{I}zzet Sakall\i{}}
\email{izzet.sakalli@emu.edu.tr}
\affiliation{Physics Department, Arts and Sciences Faculty, Eastern Mediterranean
University, Famagusta, North Cyprus via Mersin 10, Turkey.}

\author{Joel Saavedra}
\email{joel.saavedra@pucv.cl}
\affiliation{Instituto de F\'{\i}sica, Pontificia Universidad Cat\'olica de Valpara\'{\i}%
so, Casilla 4950, Valpara\'{\i}so, Chile.}
\date{\today }

\begin{abstract}
We study the quasinormal modes (QNMs) of the Schwarzschild black
hole immersed in an electromagnetic (em) universe. The immersed
Schwarzschild black hole (ISBH) is originated from the metric of colliding
em waves with double polarization [Class. Quantum Grav. 12, 3013 (1995)].
The perturbation equations of the scalar fields for the ISBH geometry are
written in the form of separable equations. We show that these equations can
be transformed to the confluent Heun's equations, for which we are able to
use the known techniques to perform the analytical quasinormal (QNM)
analysis of the solutions. Furthermore, we employ numerical methods
[Mashhoon and $6^{th}$-order Wentzel-Kramers-Brillouin (WKB)] to
derive the QNMs. The results obtained are discussed and depicted with the
appropriate plots.
\end{abstract}

\maketitle

\section{Introduction}

Classical black holes are closed systems that do not emit any signal to an
outside observer. The only way to obtain information from a black hole is to
study its relativistic wave dynamics with quantum mechanics (e.g., Hawking
radiation, QNMs, and gravitational waves). To have QNMs, a black hole must
be perturbed. A fair analogy to this concept is the ringing of a bell, which
is a damped harmonic oscillator. The perturbation of a black hole has at
least three stages: (i) the transient, which depends on the initial
perturbation; (ii) the QNM ring-down, which is an important stage that
reveals unique frequencies containing information about the source; and
(iii) the exponential/power-law tail, which occurs when the energy is very
low at the end of the perturbation. QNMs can be found by applying
perturbation to the black hole spacetime with appropriate boundary
conditions: the wave solution should be purely outgoing at infinity and
purely ingoing at the event horizon \cite{fernando1,J1,bw1,S1,cec}. For the
remarkable review and research papers of the QNMs, a reader may refer to 
\cite{rev4,rev5,rev6,rev7,rev8,rev9,rev10,rev11}. Detection of gravitational
waves \cite{DGW1,DGW2,DGW3} have brought the QNMs under the spotlight again.
On the other hand, the QNMs (having frequencies above 500 Hz \cite{500hz}) of lower mass
black holes and neutron star mergers signatures are presently not detectable. The main problem of it is the increasing quantum shot noise \cite{SHOT} at
the high frequenciy regime. However, recent developments  
\cite{GWQM} are
very promising for the detection of QNMs in the near future.

Our main aim in this study is to study the QNMs of massive/massless scalar
fields in the ISBH spacetime. To this end, we shall use particular
analytical and numerical methods. Iyer and Will \cite{Iyer} are the first
researchers, who obtained QNMs with the help of the third order WKB
approximation. Later on, their study was extended to the sixth order by
Konoplya and Zhidenko \cite{kono,kono2,kono3}. In the sequel, the WKB
approximations are considered by other researchers to compute the QNMs of
various spacetimes \cite%
{fernando2,fernando3,xi,fernando4,fernando5,J2,J3,J4,J5,J6,J7,bw2,bw22,bw3,bw4,bw5,bw6,bw7,Jansen:2017oag,S2,S3,S4,ann1,ann2,a2,a3,a4,a5}.

The ISBH solution is given by \cite{h1,h2,h3}:

\begin{equation}
ds^{2}=-F(r)dt^{2}+\frac{1}{F(r)}dr^{2}+r^{2}\left( d\theta ^{2}+sin(\theta
)^{2}d\varphi ^{2}\right) ,  \label{1}
\end{equation}%
where 
\begin{equation}
F(r)=1-\frac{2M}{r}+\frac{M^{2}(1-a^{2})}{r^{2}},  \label{2}
\end{equation}%
\textit{in which }$M$\textit{\ denotes the mass-parameter and }$a$\textit{\
is the interpolation parameter \cite{h3,a1}: }$1\geq a\geq 0$\textit{.
Letting the effective charge as } $Q_{eff}\equiv M^{2}(1-a^{2}),$ \textit{it
is clear that when }$a=1$ \textit{i.e.,} $Q_{eff}=0$\textit{, metric (1) is nothing
but the Schwarzschild black hole. However, the case of }$a=0$\textit{\ (}$%
Q_{eff}=M^{2}$\textit{) corresponds to the Reissner-Nordstr\"{o}m black hole 
\cite{RN1,RN2}. On the other hand, the metric function }$F(r)$\textit{\ can
be rewritten as}
\begin{equation}
F(r)=\frac{\left( r-r_{p}\right) \left( r-r_{n}\right) }{r^{2}},  \label{3n}
\end{equation}%
\textit{where }$r_{p}=M(1+a)$\textit{\ and }$r_{n}=M(1-a)$\textit{\ are the
event and inner horizons, respectively. To illustrate the effect of }$a$%
\textit{-parameter on the em structure of the spacetime, one
can use the Newman-Penrose formalism \cite{mych}. To this end, the null
tetrad frame }$\left( l,n,m,\overline{m}\right) $\textit{, which satisfies
the orthogonality conditions (}$l.n=-m$\textit{.}$\overline{m}=1$\textit{)
are choosen to be}%
\begin{equation*}
l_{\mu }=dt-\frac{dr}{F\left( r\right) },
\end{equation*}%
\begin{equation*}
2n_{\mu }=F\left( r\right) dt+dr,
\end{equation*}%
\ 
\begin{equation*}
\sqrt{2}m_{\mu }=-r(d\theta +i\sin \theta d\phi ),
\end{equation*}

\begin{equation}
\sqrt{2}\overline{m}_{\mu }=-r(d\theta -i\sin \theta d\phi ).
\end{equation}

\textit{Thus, the non-zero Weyl and Ricci scalars can be computed as follows}
\begin{equation}
\Psi _{2}=-\frac{(r_{p}+r_{n})r-2r_{p}r_{n}}{2r^{4}}=-\frac{1}{r^{4}}\left[
Mr+Q_{eff})\right] ,  \label{Weyl}
\end{equation}%
\begin{equation}
\Phi _{11}=\frac{r_{p}r_{n}}{2r^{4}}=\frac{Q_{eff}}{2r^{4}}.
\end{equation}%

\textit{Since the only non-zero Weyl scalar is Eq. (\ref{Weyl}), the metric (%
\ref{1}) represents a Petrov type-D [xx] spacetime. The effect of }$a$%
\textit{-parameter on gravitation and em fields is now more
clear:}

\begin{equation}
\textit{Gravity}\rightarrow \left( a=1\right) \rightarrow \Psi _{2}=\frac{%
-M}{r^{3}},\text{ }\Phi _{11}=0.  \notag
\end{equation}

\begin{equation*}
\textit{Gravity + em}\rightarrow \left( 0\leq a<1\right)
\rightarrow \Psi _{2}\neq 0\neq \Phi _{11}.
\end{equation*}

The Hawking temperature \cite{SAKL0} is expressed in terms of the surface
gravity ($\kappa $) as $T_{H}=\frac{\kappa }{2\pi }.$ For the ISBH geometry,
it is given by
\begin{equation}
T_{H}=\frac{\kappa }{2\pi }=\left. \frac{F^{\prime }(r)}{4\pi }\right\vert
_{r=r_{p}}=\frac{a}{2M\pi(a+1)^{2}}.  \label{4n}
\end{equation}

This paper is organized as follows: in Section II, we provide a complete
analytical solution to the massive Klein-Gordon equation (KGE) in terms of
the confluent Heun functions. We then show how the QNMs can be computed from
that obtained exact solution. Sections III and IV are devoted to the
numerical studies of the massless KGE in the ISBH geometry. We obtain the
corresponding effective potential and analyze it, thoroughly. We present two
numerical methods (the Mashhoon and the sixth-order WKB) for computing the
QNMs of the ISBH. Finally, we summarize our discussions in the conclusion.

\section{Analytical QNMs}

Let us consider a massive scalar field that obeys the KGE on the ISBH metric
(1). Recall that a massive KGE is given by (e.g., \cite{SAKL1})

\begin{equation}
\frac{1}{\sqrt{-g}}\partial _{\alpha }\left( \sqrt{-g}g^{\alpha \beta
}\partial _{\beta }\Psi _{0}\right) -\mu _{0}^{2}\Psi _{0}=0.  \label{1sn}
\end{equation}
Here, $\mu _{0}$ and $\Psi _{0}$\ represent the mass and the scalar field,
respectively. It is straight forward to see that Eq. (\ref{1sn}) is
separable with the following ansatz: 
\begin{equation}
\Psi _{0}=\Psi _{0}(\boldsymbol{r},t)=\mathcal{R}(r)\mathcal{A}(\theta
)e^{im\varphi }e^{-i\omega t},  \label{2sn}
\end{equation}%
where $\omega $ denotes the frequency of the wave and $m$ denotes the
magnetic quantum number associated with the rotation in the $\varphi $
direction. By defining an eigenvalue ($\lambda $), one can show that Eq. (%
\ref{1sn}) leads to the following angular and radial equations:

\begin{equation}
\sin \left( \theta \right) {\frac{d^{2}}{d{\theta }^{2}}}\mathcal{A}\left(
\theta \right) +\left( {\frac{d}{d\theta }}\mathcal{A}\left( \theta \right)
\right) \cos \left( \theta \right) +\left( \lambda \sin \left( \theta
\right) -{\frac{{m}^{2}}{\sin \left( \theta \right) }}\right) \mathcal{A}%
\left( \theta \right) =0,  \label{3sn}
\end{equation}

and

\begin{equation}
\left( {\frac{d}{dr}}\Delta \left( r\right) \right) {\frac{d}{dr}}\mathcal{R}%
\left( r\right) +\Delta \left( r\right) {\frac{d^{2}}{d{r}^{2}}}\mathcal{R}%
\left( r\right) +\left( {\frac{{r}^{4}{\omega }^{2}}{\Delta \left( r\right) }%
}-\lambda -\mu _{0}^{2}{r}^{2}\right) \mathcal{R}\left( r\right) =0.
\label{4sn}
\end{equation}

The solution of the angular equation (\ref{3sn}) is given in terms of the
four-dimensional spheroidal functions \cite{SAKL2}. To obtain the exact
solution of the radial equation (\ref{4sn}), we first introduce a new
variable: 
\begin{equation}
z=\left(r-r_{p}\right) \Bbbk _{m}^{-1},  \label{5sn}
\end{equation}%
where $\Bbbk _{m}=r_{p}-r_{n}$. By using Eq. (\ref{5sn}) in Eq. (\ref{4sn}),
one can transform the radial equation into the following form: 
\begin{equation}
z\left( 1-z\right) {\frac{d^{2}}{d{z}^{2}}}\mathcal{R}\left( z\right)
+\left( 1-2z\right) {\frac{d}{dz}}\mathcal{R}\left( z\right) +\left( \lambda
+\mu _{0}^{2}\left( r_{p}-\Bbbk _{m}z\right) ^{2}-{\frac{\left( r_{p}-\Bbbk
_{m}z\right) ^{4}{\omega }^{2}}{z\left( z-1\right) \Bbbk _{m}^{2}}}\right) 
\mathcal{R}\left( z\right) =0.  \label{6sn}
\end{equation}

Furthermore, applying a particular s-homotopic transformation \cite{SAKL3}
to $\mathcal{R}(z)$: 
\begin{equation}
\mathcal{R}(z)=\mbox{e}^{B_{1}z}z^{B_{2}}(1-z)^{B_{3}}\mathcal{U}(z),
\label{7sn}
\end{equation}%
where the coefficients $B_{1}$, $B_{2}$, and $B_{3}$ are given by 
\begin{equation}
B_{1}=\Bbbk _{m}\sqrt{\mu _{0}^{2}-{\omega }^{2}},  \label{8sn}
\end{equation}%
\begin{equation}
B_{2}={\frac{ir_{p}^{2}\omega }{\Bbbk _{m}}},  \label{9sn}
\end{equation}%
\begin{equation}
B_{3}={\frac{ir_{n}^{2}\omega }{\Bbbk _{m}}},  \label{10sn}
\end{equation}

we obtain a differential equation for $\mathcal{U}(z)$, which is identical
to the confluent Heun differential equation \cite%
{SAKL4,SAKL5,SAKL6,SAKL7,SAKL8,SAKL82,BIRK,SAKL9} (for one of the most
detailed works about the applications of the Heun differential equation, the
reader is referred to \cite{Kraniotis}):

\begin{equation}
{\frac{d^{2}}{d{z}^{2}}}\mathcal{U}(z)+\left( \widetilde{a}+\frac{1+%
\widetilde{b}}{z}-\frac{1+\widetilde{c}}{1-z}\right) {\frac{d}{dz}}\mathcal{U%
}(z)+\left( \frac{\widetilde{f}}{z}-\frac{\widetilde{g}}{1-z}\right) 
\mathcal{U}(z)=0.  \label{11sn}
\end{equation}

The three parameters seen in the coefficient bracket of ${\frac{d}{dz}}%
\mathcal{U}(z)$ are given by

\begin{equation}
\widetilde{a}=2B_{1},\ \ \widetilde{b}=2B_{2},\ \ \widetilde{c}=2B_{3}.
\label{12sn}
\end{equation}

Setting

\begin{equation}
\widetilde{d}=-\Bbbk _{m}\Bbbk _{p}\left( \mu _{0}^{2}-2\,{\omega }%
^{2}\right),  \label{13sn}
\end{equation}

\begin{equation}
\widetilde{e}={\frac{-r_{p}^{2}\left[ \left( \mu _{0}\Bbbk _{m}\right)
^{2}-2r_{p}\,{\omega }^{2}\left( \Bbbk _{m}-r_{n}\right) \right] -\Bbbk
_{m}^{2}\lambda }{\Bbbk _{m}^{2}}},  \label{14sn}
\end{equation}

where $\Bbbk _{p}=r_{p}+r_{n}$, one can also find the other two parameters
of Eq. (\ref{11sn}) as follows:

\begin{equation}
\widetilde{f}=\frac{1}{2}(\widetilde{a}-\widetilde{b}-\widetilde{c}+%
\widetilde{a}\widetilde{b}-\widetilde{b}\widetilde{c})-\widetilde{e},
\label{15sn}
\end{equation}

\begin{equation}
\widetilde{g}=\frac{1}{2}(\widetilde{a}+\widetilde{b}+\widetilde{c}+%
\widetilde{a}\widetilde{c}+\widetilde{b}\widetilde{c})+\widetilde{d}+%
\widetilde{e}.  \label{16sn}
\end{equation}

The general solution of the confluent Heun differential equation (\ref{11sn}%
) is given by \cite{SAKL10} as follows

\begin{equation}
\mathcal{U}(z)=c_{1}\mbox{HeunC}(\widetilde{a},\widetilde{b},\widetilde{c},%
\widetilde{d},\widetilde{e};z)+c_{2}z^{-\widetilde{b}}\mbox{HeunC}(%
\widetilde{a},-\widetilde{b},\widetilde{c},\widetilde{d},\widetilde{e};z),
\label{17sn}
\end{equation}
where $c_{1}$ and $c_{2}$ are integration constants. Thus, the general
solution of Eq. (\ref{6sn}) in the exterior region of the event horizon ($%
0\leq z<\infty $) reads 
\begin{equation}
\mathcal{R}(z)=\mbox{e}^{B_{1}z}(1-z)^{B_{3}}\Bigg[C_{1}z^{B_{2}}\mbox{HeunC}%
(\widetilde{a},\widetilde{b},\widetilde{c},\widetilde{d},\widetilde{e}%
;z)+C_{2}z^{-B_{2}}\mbox{HeunC}(\widetilde{a},-\widetilde{b},\widetilde{c},%
\widetilde{d},\widetilde{e};z)\Bigg].  \label{18sn}
\end{equation}

Now, we follow one of the recent techniques \cite{SAKL7,SAKL9,SAKL11} to
compute the QNMs of scalar waves propagating in the geometry of an ISBH. As
is well known, QNMs are the solutions associated to those complex
frequencies. In particular, the imaginary component of the frequency states
how fast the oscillation will fade over time \cite{SAKL13}.

The QNMs can be obtained from the radial solution (\ref{18sn}) under certain
boundary conditions: the Heun functions should be well-behaved at spatial
infinity and finite on the horizon. This requires $\mathcal{R}(z)$ to take
the form of Heun's polynomials \cite{SAKL14}, which is possible with the $%
\widetilde{\delta }_{n}$ condition \cite{SAKL9,SAKL11,SAKL00} 
\begin{equation}
\frac{\widetilde{d}}{\widetilde{a}}+\frac{\widetilde{b}+\widetilde{c}}{2}%
+1=-n\ ,\text{ \ \ \ \ \ \ \ \ with \ \ \ }n=0,1,2,\ldots \   \label{19sn}
\end{equation}

In \cite{SAKL7}, it is shown that the Heun's polynomials arising from Eq. (%
\ref{19sn}) yield the most general class of solutions to the Teukolsky
master equation pertinent to the Teukolsky-Starobinsky identities \cite%
{SAKL15}, which are closely related to the subject of QNMs \cite%
{SAKL16,SAKL17}. Using Eq. (\ref{19sn}), we find out that assuming ${\omega
\geq }\mu _{0}$

\begin{equation}
i\left[ {\frac{-\Bbbk _{p}\left( 2\,{\omega }^{2}-\mu _{0}^{2}\right) }{2%
\sqrt{{\omega }^{2}-\mu _{0}^{2}}}}+{\frac{\omega \,\left(
r_{p}^{2}+r_{n}^{2}\right) }{\Bbbk _{m}}}\right] +1=-n .  \label{20sn}
\end{equation}

With the aid of a mathematical computer package like Maple 18 \cite{SAKL10},
one can obtain a solution for ${\omega }$ from Eq. (\ref{20sn}). However,
the solution is excessively lengthy, which prevents us from typing it here.
On the other hand, if one considers the very light spin-0 particles having a 
$\mu _{0}\sim 0$, the $\widetilde{\delta }_{n}$ condition (\ref{19sn})
results in

\begin{equation}
\frac{2\,i\omega \,r_{p}^{2}}{\Bbbk _{m}}+1=-n.  \label{21sn}
\end{equation}

The above equation allows the following solution for the QNMs:

\begin{equation}
\omega _{n}=i{\frac{\Bbbk _{m}}{2r_{p}^{2}}}\left( n+1\right).  \label{22sn}
\end{equation}

Recalling the definition of surface gravity ($\kappa $) from Eq. (\ref{4n}),
we have

\begin{equation}
\kappa =\frac{\Bbbk _{m}}{2r_{p}^{2}}.  \label{23sn}
\end{equation}

This changes Eq. (\ref{22sn}) to the following form:

\begin{equation}
\omega _{n}=i\kappa \left( n+1\right)= i2\pi T_H\left( n+1\right).
\label{24sn}
\end{equation}

It is obvious from Eq. (\ref{24sn}) that ISBH admits pure imaginary QNMs,
which are in good agreement with the QNM result for the Schwarzschild black
hole with a global monopole \cite{SAKL00}. On the other hand, although it is
valid for $n\rightarrow\infty$, the QNM result of Hod \cite{SAKL18} (see
also \cite{SAKL19}), which was analytically obtained by the
continued-fraction argument method \cite{SAKL20} supports also our result (%
\ref{24sn}).

\section{QNMs by the Mashhoon method: Approximation with Poschl-Teller
potential}

As we have experience from the previous section, the mass does not play an
important role on the QNMs. Because of this fact, we consider the following
massless KGE to perform numerical analysis in this section: 
\begin{equation}
\nabla ^{2}\varPhi=0\; .  \label{5}
\end{equation}

\begin{figure}[!htb]
\centering
\includegraphics[width=3 in]{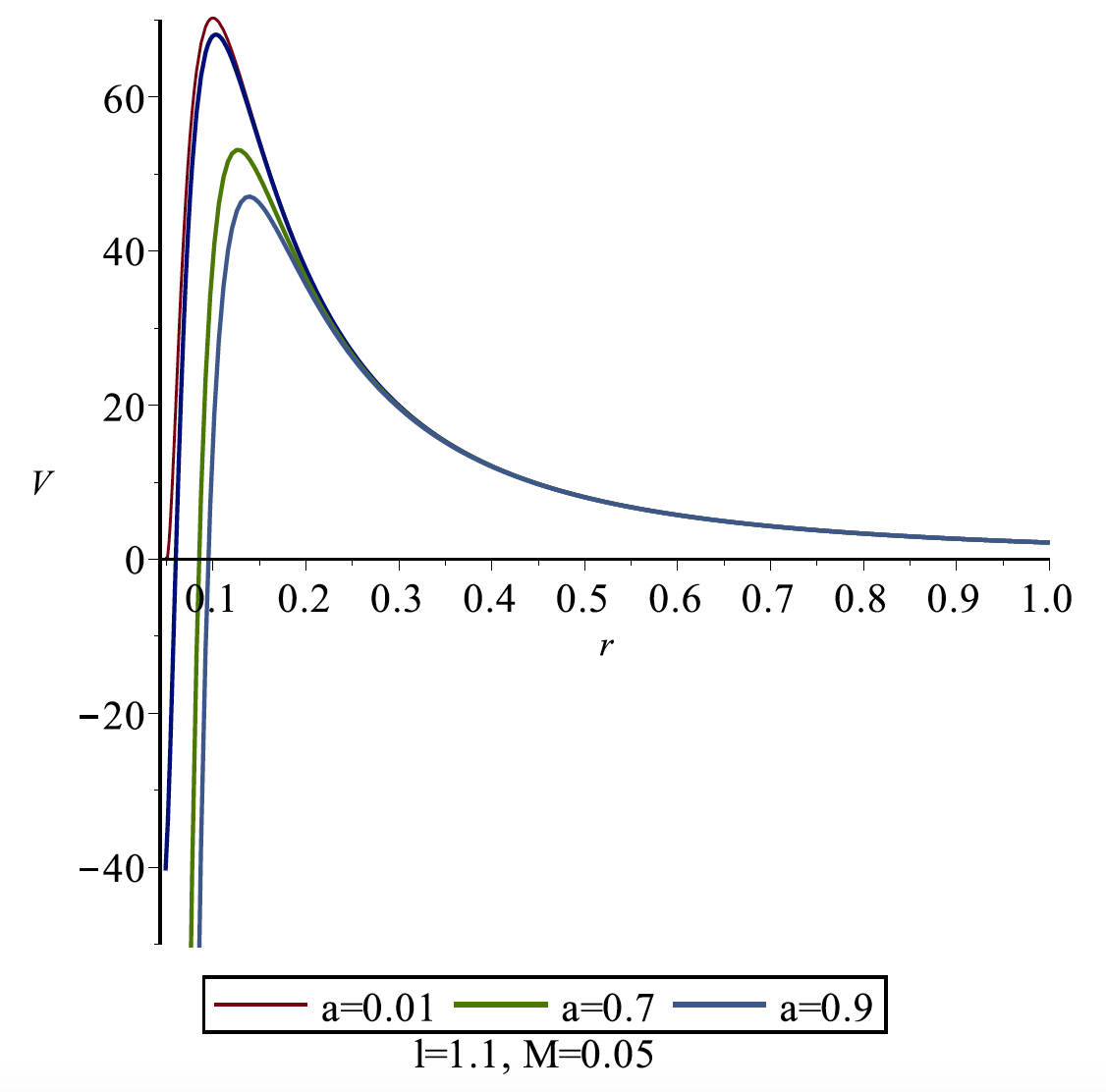}
\caption{The figure shows $\mathit{V}$ vs. $r$ for massless particles.}
\label{Figpot}
\end{figure}

By taking the ansatz of the scalar field, which can be decomposed into its
partial modes in terms of the spherical harmonics $Y_{l,m}(\theta ,\varphi )$%
, 
\begin{equation}
\varPhi(r,\theta ,\phi ,t)=\frac{R(r)}{r}Y_{lm}(\theta ,\phi )e^{-i\omega t}.
\label{6}
\end{equation}%
Here, $\omega ,$ $l,$ and $m$ are the oscillating frequency of the scalar
field, the angular quantum number, and the magnetic quantum number,
respectively. Then, we separate the massless KGE to obtain the following
radial differential equation:

\begin{equation}
R^{\prime \prime }+\left( \omega ^{2}-\mathit{{V}_{0}(r)}\right) R=0,
\label{7wave}
\end{equation}

where the effective potential is given by 
\begin{equation}
\mathit{{V}_{0}(r)=F(r)\left[ \frac{\left( l(l+1\right) }{r^{2}}+\frac{%
F^{\prime }}{r}\right].}  \label{8}
\end{equation}%
Note that a prime stands for the derivate with respect to the tortoise
coordinates $(r_{\ast })$, $dr_{\ast }=\frac{dr}{F(r)}$. First, we investigate the features of the potential by plotting it with different values of parameters such as $M$ and $a$. In Fig. \ref{Figpot}, the potential is plotted for various values of $a$. It is obvious that when $a$ increases, the height decreases.

Now, we use the Mashhoon method to calculate the QNMs, numerically \cite%
{kono3-1,mash,grtensor}. Wave functions vanish at the boundaries and the QNM
problem becomes a bound-states problem with a potential of $\mathit{{V}%
_{0}\rightarrow -{V}_{0}}$. Moreover, analytic solutions of the wave
equation for this kind of potential resemble the Poschl-Teller (PT)
potential 
\begin{equation}
V_{PT}=\frac{\mathit{{V}_{0_{max}}}}{cosh^{2}\alpha (r_{\ast }-r)}.
\label{14n}
\end{equation}

Here, $\mathit{{V}_{0_{max}}}$ is the effective potential (\ref{8}) at the
maximum point, which gives the height. The bound states of the PT potential
are portrayed as follows: 
\begin{equation}
\omega (\alpha )=W(\alpha ^{\prime }),  \label{15n}
\end{equation}

\begin{equation}
W=\alpha ^{\prime }\left[ -(n+\frac{1}{2})+\frac{1}{4}+\sqrt{\frac{\mathit{{V%
}_{0_{max}}}}{\alpha ^{\prime 2}}}\right].  \label{16n}
\end{equation}

The QNMs $(\omega)$ are calculated using the inverse of the PT potential
bound states ($\alpha ^{\prime }=i\alpha $). Thus, we have \cite%
{kono3-1,mash}

\begin{equation}
\omega =\pm \sqrt{\mathit{{V}_{0_{max}}-\frac{1}{4}\alpha ^{2}}}-i\alpha (n+%
\frac{1}{2}),  \label{17n}
\end{equation}%
where $n$ is the overtone number, and $\omega $ is calculated for varying
values of $n$: $(-1.5, i 0.3230265022)$, $(-3.0, i 0.9243084555)$, $(-7.5, i
2.470697496)$, and $(-15.0, i 4.985413335)$. It is clear that the field
decays faster for large values of $a$. From the above solution it is seen that the perturbations are stable ( $%
Im\omega < 0$) as well as the damping increases with the overtone number $n$.

\section{Numerical Results with the sixth-order order WKB method}

In this section, by employing the Konoplya's sixth order WKB approach \cite%
{kono}, we compute the QNM frequencies and obtain the QNMs from the
following identity:

\begin{equation}
\frac{\omega ^{2}-V_{0}}{\sqrt{-2V^{\prime \prime }_{0}}}%
-L_{2}-L_{3}-L_{4}-L_{5}-L_{6}=\left( n+\frac{1}{2}\right).  \label{18n}
\end{equation}

Here, $\mathit{{V}^{\prime \prime }_{0_{max}}}$ is the second derivative of
the maximum effective potential. Details of the expressions for $L_{i}$ can
be found in \cite{kono}, and $n$ is the overtone number. The last value is
the maximum point of the potential.

The QNM frequencies are given by $\omega =\omega _{R}-i\omega _{I}$. A
positive imaginary value of $i\omega _{I}$ means that it is damped and
negative $i\omega _{I}$ means that there is an instability.

The result of Eq. (\ref{18n}) admits the list of QNMs found with sixth-,
fifth-, fourth-, third-, and second-order WKB expressions and the eikonal
approximations with different values of multipole number $l=1, 2, 3, 4$ and $%
a=0.1, 0.5, 0.9$:

\begin{longtable}[c]{|l|l|l|l|l|l|l|}
\hline
$l$ & $\omega_1$  & $\omega_2$  & $\omega_3$  & $\omega_4$  & $\omega_5$  & $\omega_6$  \\ \hline
\endfirsthead
\endhead

1 & 0.376627-i0.089963 & 0.376291-i0.0900433 &0.376188-i0.0896098& 0.374267-i0.0900696 & 0.376185-i0.0977338  & 0.4041-i0.0909823\\ \hline
2 & 0.623756-i0.08954 & 0.623774-i0.0895374&0.623761-i0.0894481& 0.623273-i0.0895182& 0.623696-i0.0924193  & 0.640826-i0.0899488 \\ \hline
3 & 0.872001-i0.0893856 & 0.872008-i0.0893849 & 0.872006-i0.089359 & 0.871821-i0.089378& 0.871975-i0.0908726& 0.884262-i0.0896099\\ \hline
4 & 1.12049-i0.0893221  & 1.12049-i0.089322 & 1.12049-i0.0893121  & 1.1204-i0.0893192& 1.12047-i0.0902263 & 1.13004-i0.0894622  \\ \hline
\caption{a=0.1}
\label{tablea01}\\
\end{longtable}

\begin{longtable}[c]{|l|l|l|l|l|l|l|}
\hline
$l$ & $\omega_1$  & $\omega_2$  & $\omega_3$  & $\omega_4$  & $\omega_5$  & $\omega_6$  \\ \hline
\endfirsthead
\endhead
1 & 0.345998-i0.0981664 & 0.34612-i0.0981318& 0.346002-i0.0977128& 0.344377 - i0.0981738& 0.346651-i0.105874 & 0.377539-i0.0972126\\ \hline
2 & 0.570878-i0.0975128 & 0.5709-i0.097509& 0.570887-i0.0974313& 0.570478-i0.097501& 0.570995-i0.100477 & 0.590496-i0.0971589 \\ \hline
3 & 0.797048-i0.0973316 & 0.797053-i0.097331 & 0.79705-i0.0973092 & 0.796896-i0.097328& 0.797086-i0.0988706& 0.811203-i0.0971499 \\ \hline
4 & 1.02363-i0.0972581  & 1.02363-i0.0972579  & 1.02363-i0.0972497  & 1.02355-i0.0972567 & 1.02364-i0.0981953 & 1.03468-i0.0971477  \\ \hline
\caption{a=0.5}
\label{tablea05}\\
\end{longtable}

\begin{longtable}[c]{|l|l|l|l|l|l|l|}
\hline
$l$ & $\omega_1$  & $\omega_2$  & $\omega_3$  & $\omega_4$  & $\omega_5$  & $\omega_6$  \\ \hline
\endfirsthead
\endhead

1 & 0.303024-i0.0986391& 0.303153-i0.0985971&0.303055-i0.0982969& 0.301332-i0.0988591& 0.338945-i0.0971429& 0.338945-i0.0971429\\ \hline
2 & 0.500192-i0.097696 &0.500207-i0.097693 & 0.500198-i0.0976455& 0.499781-i0.097727& 0.500499-i0.101335& 0.522507-i0.097067 \\ \hline
3 & 0.698423-i0.0974405 & 0.698426-i0.0974402& 0.698424-i0.097428& 0.698269-i0.0974496& 0.698533-i0.0993224&  0.714528-i0.0970991 \\ \hline
4 & 0.897001-i0.097336 & 0.897002-i0.0973359 &0.897001-i0.0973315  &0.896928-i0.0973395  &0.897052-i0.0984799&0.909582-i0.0971233  \\ \hline
\caption{a=0.9}
\label{tablea09}\\
\end{longtable}

\begin{figure}[!htb]
\centering
\includegraphics[width=3 in]{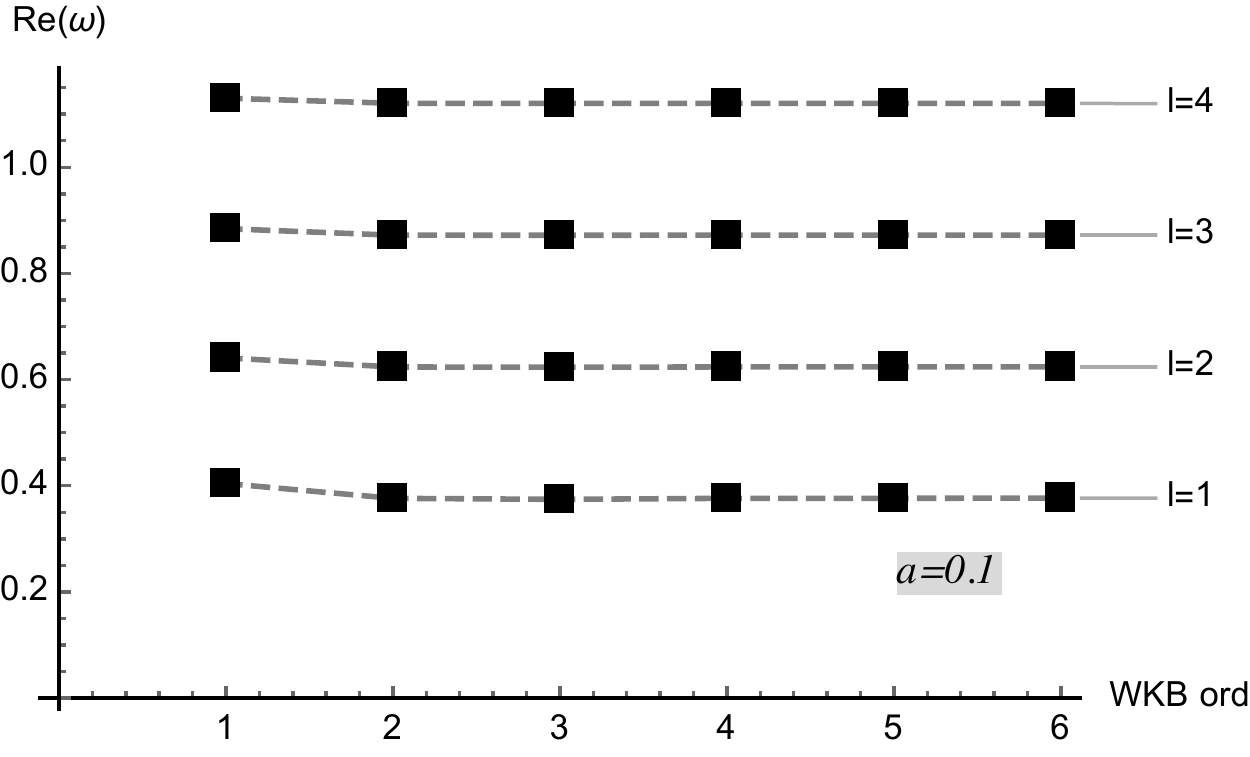} 
\includegraphics[width=3 in]{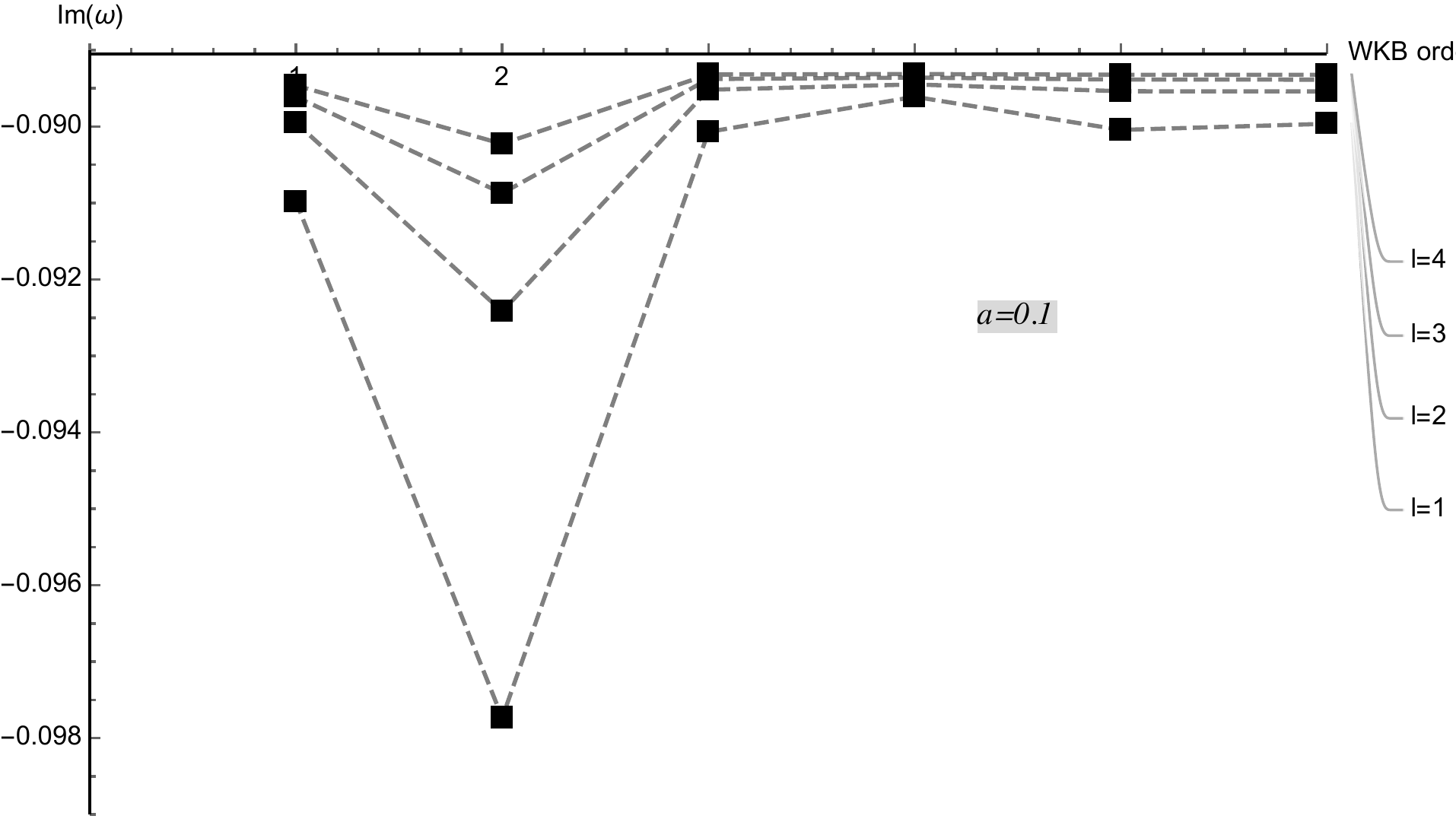}
\caption{The real and imaginary part of QNMs with sixth-, fifth-, fourth-,
third-, and second-order WKB formula and the eikonal approximation for $%
a=0.1 $, $M=1$ and $s=0$ mode with the multipole numbers $l=(1, 2, 3, 4)$.}
\label{plot2}
\end{figure}

\begin{figure}[!htb]
\centering
\includegraphics[width=3 in]{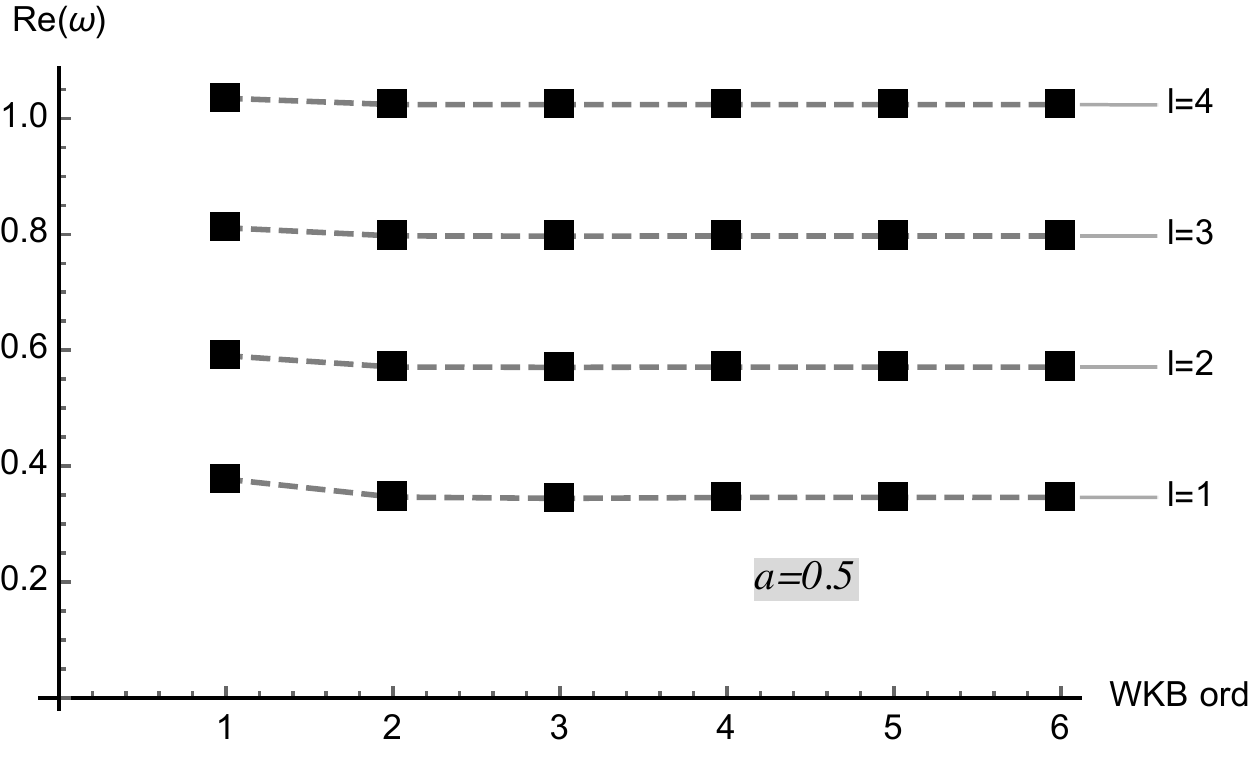} 
\includegraphics[width=3in]{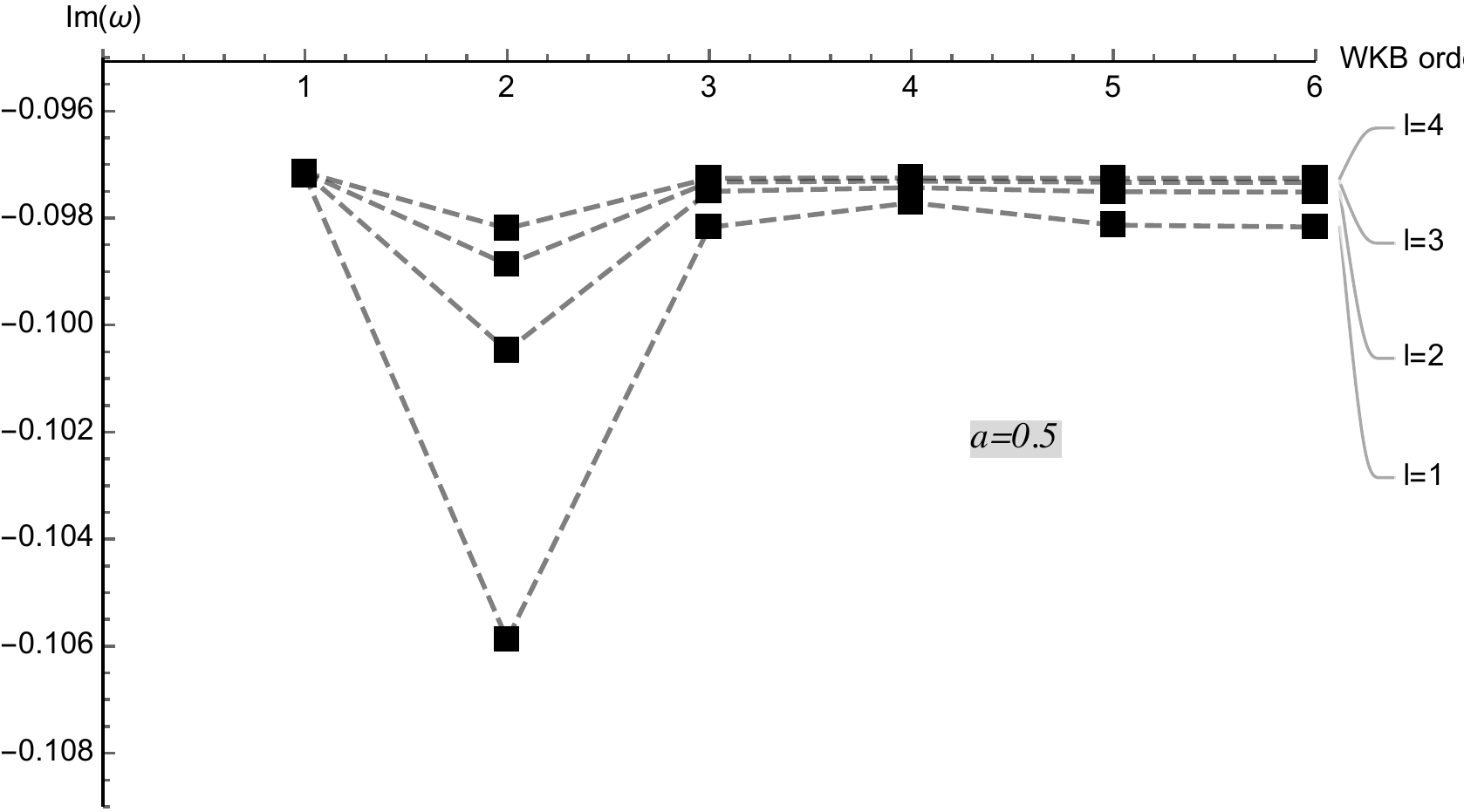}
\caption{The real and imaginary part of QNMs with sixth-, fifth-, fourth-,
third-, and second-order WKB formula and the eikonal approximation for $%
a=0.5 $, $M=1$ and $s=0$ mode with the multipole numbers $l=(1, 2, 3, 4)$.}
\label{plot3}
\end{figure}

\begin{figure}[!htb]
\centering
\includegraphics[width=3 in]{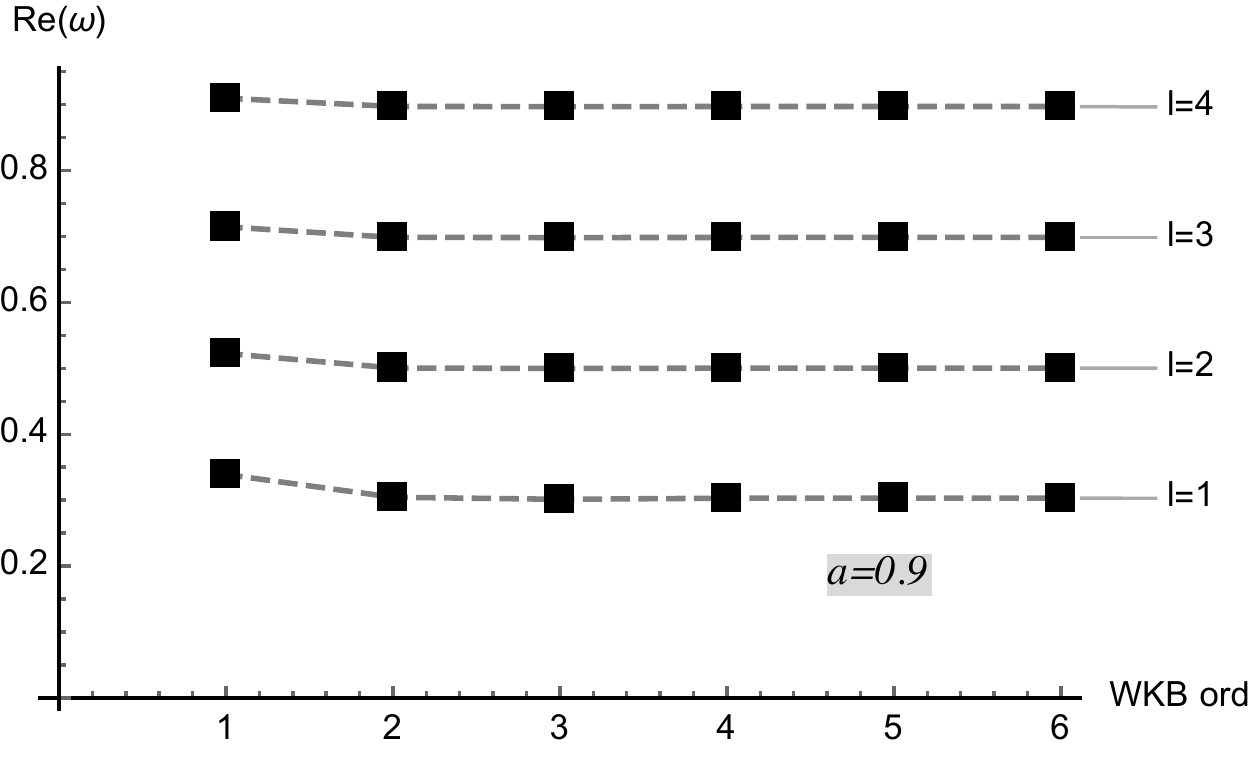} 
\includegraphics[width=3in]{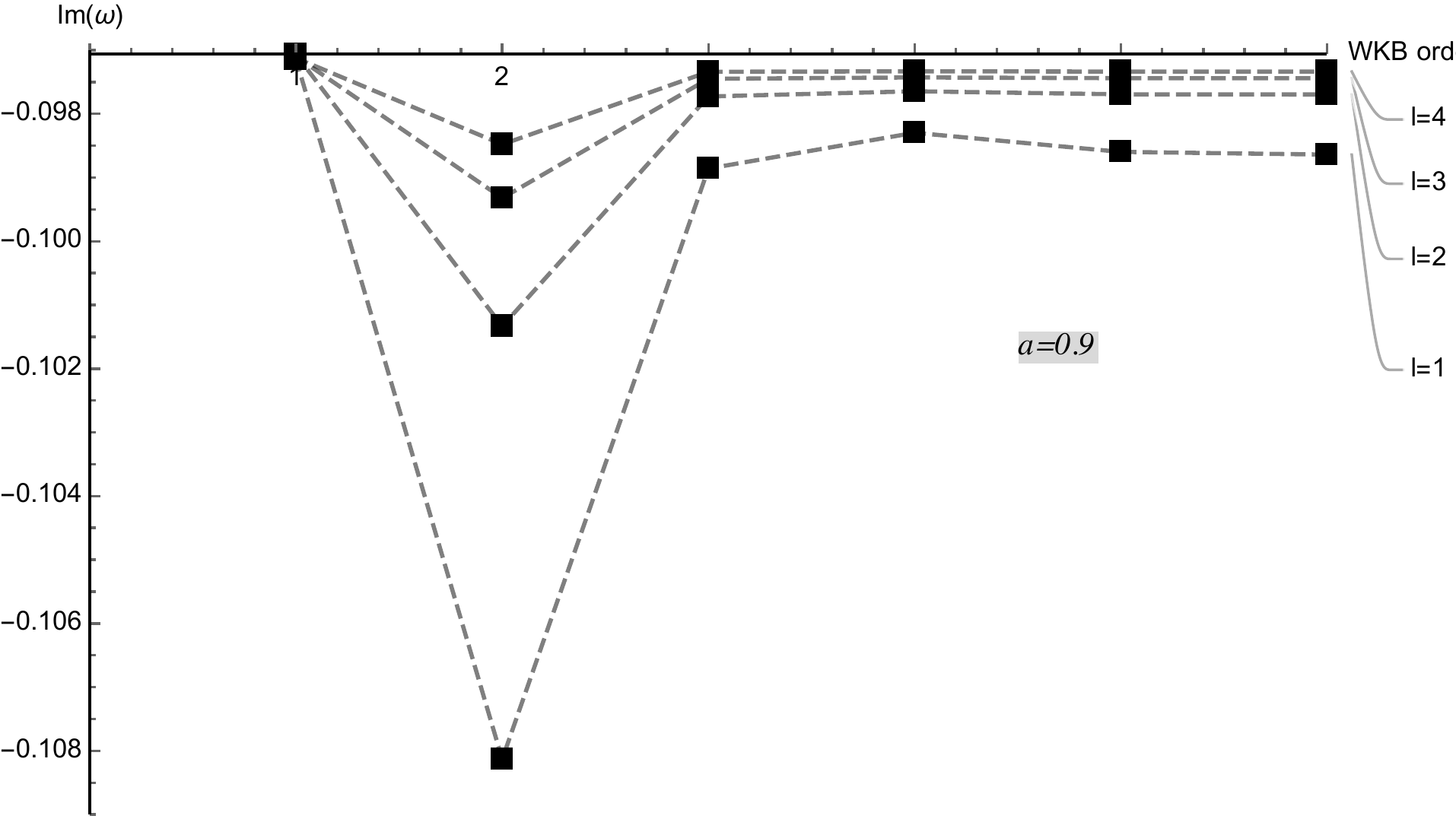}
\caption{The real and imaginary part of QNMs with sixth-, fifth-, fourth-,
third-, and second-order WKB formula and the eikonal approximation for $%
a=0.9 $, $M=1$ and $s=0$ mode with the multipole numbers $l=(1, 2, 3, 4)$.}
\label{plot4}
\end{figure}

The convergence of the WKB formula for varying values of $a$ and the
expedited field decay can be seen in Tables (\ref{tablea01}, \ref{tablea05}, %
\ref{tablea09}) and plotted in Figs. (\ref{plot2}, \ref{plot3}, \ref{plot4})
for values of $a=(0.1, 0.5, 0.9)$, $M=1$ and $s=0$ mode with the multipole
numbers $l=(1, 2, 3, 4)$.

\begin{figure}[!htb]
\centering
\includegraphics[width=3 in]{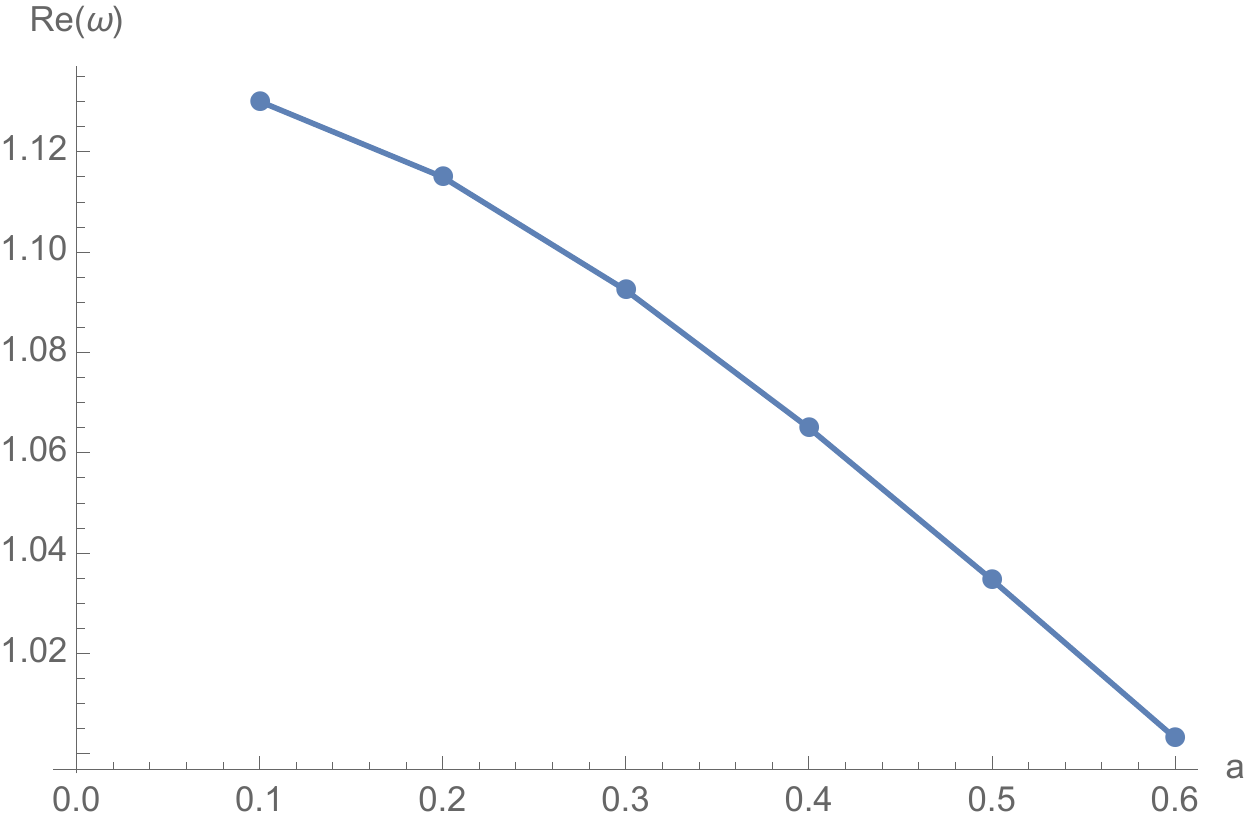} 
\includegraphics[width=3 in]{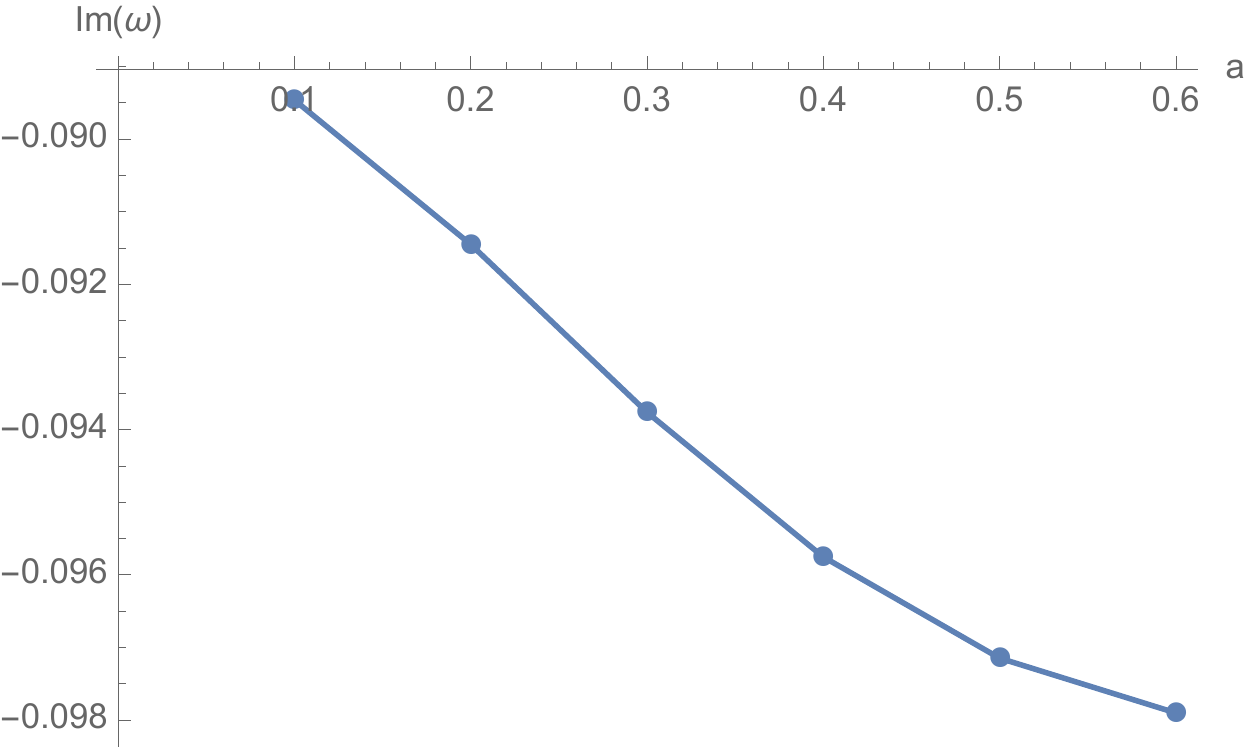}
\caption{The real and imaginary part of QNMs with sixth-order WKB formula for $a$ versus $Re(\omega)$ and $a$ versus $Im(\omega)$ for $M=1$, $s=0$ mode with the multipole number $l=4$.}
\label{plot5}
\end{figure}

\section{Conclusion}

In this paper, we have analytically studied massive scalar field
perturbations by using the KGE in the geometry of ISBH. After finding the
exact solution of the radial wave equation, we have found close agreement
with the obtained QNMs of the ISBH and previous studies \cite%
{SAKL18,SAKL19,SAKL20}, which were about QNMs of Schwarzschild black hole.

We have then employed the sixth-order WKB approximation method to compute
the QNMs. Special attention has been paid to the details of how QNMs vary
with the interpolation parameter $a$. The plots of the QNM frequencies $%
\omega $ versus $a$ show that the real part of the QNM frequency $Re\omega $
decreases with $a$, similarly to the imaginary part of the QNM frequency $Im\omega $
decreases with it shown in Fig. \ref{plot5}. We have also inferred from the associated graphs that if
one plots the QNM frequencies from the lower to the higher overtones, taking
into account different WKB orders, the comparative accuracy gets better when 
$l<n$. Namely, similar to the study of QNMs of test
fields around regular black holes \cite{PRD2015}, the outputs [Tables (\ref{tablea01},\ref{tablea05},\ref{tablea09}) and Figs. (\ref{plot2},\ref{plot3},\ref{plot4},\ref{plot5})] have
shown that an increase of $Q_{eff}$ (i.e, $a\rightarrow0$) implies a monotonic increase
of $Re\omega$ and $Im\omega$ (and vice versa): the damping rate of the wave decreases with increasing $Q_{eff}$.
\textit{One can infer from the latter results that ISBHs' oscillators are "better"
(slowly damped) than the Schwarzschild BH. On the other hand, our results
are contrary to the QNM results obtained for the black holes in the
braneworld whose the real oscillations decreases while the damping rate
increases with increasing tidal charge parameter \cite{PRD2016}}. 

We plan to extend our study to the rotating ISBH \cite{h3} in the near future. Moreover, in addition to the scalar perturbations, we aim to study the
Dirac and Proca perturbations and  quantum tunneling processes of the ISBH. Also, the QNM frequencies of the ISBH in the eikonal limit ($l>>1$) by using the parameters of null geodesics \cite{ek1,ek2,ek3} are in our work agenda.

\acknowledgements
We are thankful to the Editor and anonymous Referees for their constructive
suggestions and comments. This work was supported by the Chilean FONDECYT
Grant No. 3170035 (A. \"{O}.). A. \"{O}. would like to thank Prof. Leonard
Susskind and Stanford Institute for Theoretical Physics for hospitality. A. 
\"{O}. is grateful to Prof. Douglas Singleton for hosting him as a research
visitor at the California State University, Fresno. Moreover, A. \"{O}. is  grateful to Institute for Advanced Study, Princeton for hospitality.

\end{document}